\documentstyle [11pt] {article}
\begin{document}
\large
\title{\bf A Lattice Boltzmann Subgrid  Model for High Reynolds Number Flows}
\author{S. Hou, J. Sterling, S. Chen and G. D. Doolen,\\
  [.5cm]{Theoretical Division}\\
        {and}\\
        {Center for Nonlinear Studies }\\
        {Los Alamos National Laboratory}\\
        {Los Alamos, NM 87545}\\[.4cm]}

\date{}
\maketitle

\vspace{0.4in}
\begin{center}
{\bf Abstract}
\end{center}

 A subgrid  turbulence model for the lattice Boltzmann method is proposed for
 high Reynolds number fluid flow applications. The method, based on the
standard
 Smagorinsky subgrid model and a single-time relaxation lattice Boltzmann
  method, incorporates the advantages of the lattice Boltzmann method for
  handling arbitrary boundaries and is easily implemented on
parallel machines. The method is applied to
  a two-dimensional driven cavity flow for studying dynamics and
the Reynolds number dependence of the flow
  structures. The substitution of  other subgrid models,
such as the dynamic subgrid model,
in the framework of the LB method is discussed.

\vspace{0.3in}

\pagebreak
\section{Introduction}

 The lattice Boltzmann (LB)
 method\cite{macn-zan,hig-jim,ccmm,qian,gary0,phy-report,schen4},
a derivative of lattice gas automaton method\cite{fhp1,wolf}, has been
successfully demonstrated to be an alternative numerical scheme to
traditional numerical methods for solving partial
 differential equations and modeling physical systems, particularly for
 simulating fluid flows with the Navier-Stokes
 equations. In traditional numerical methods, a
 given set of macroscopic equations are solved
by some specific numerical discretization. In contrast,
the fundamental principle of LB method is to
 construct a simplified molecular dynamics that incorporates the essential
 characteristics of the physical microscopic processes
so that the macroscopic averaged properties obey the desired
macroscopic equations. This microscopic approach in the LB method
incorporates several advantages of kinetic theory. It includes clear physical
pictures, easy implementation of boundaries and fully
parallel algorithms.  In particular, the LB method has been
successfully applied to problems which are usually difficult for traditional
numerical schemes, such as fluid flows through porous
media\cite{roth1,schen1}, multiphase fluid
flows\cite{roth2,somers,schen3,gun,xiaowen,daryl} and suspension motions
in fluids\cite{ladd,schen4}.

The LB method so far has been used only as a direct numerical
simulation method, {\it i.e.}, the full Navier-Stokes equations are directly
solved
without any {\em ad hoc} assumptions for constitutive relation
between turbulence stress tensor and the mean strain tensor.  Therefore,
the smallest captured scale in the LB method is the lattice unit and
the largest scale depends on the characteristic length scale in
simulation. These scales are often determined by the available computer memory.
The
LB method is consequently able to resolve relatively low Reynolds number flows.
It should be noted that although the LB method greatly reduces
noise of a system compared with lattice gas schemes, the lattice
Boltzmann method can become numerically unstable in contrast with lattice
gas methods, where the scheme is unconditionally stable.
 Numerical studies have shown that the using of the LB method
for high Reynolds number flows, without modeling  unresolved small scale
effects on
large scale dynamics, results in numerical
instability. This result is consistent with the recent argument made by
Sterling and Chen\cite{sterling} that the LB method can be viewed as an
explicit  second-order finite-difference discretization scheme.
The objective of this paper is to provide a numerical method
based on the LB method which can simulate fluid flows at high
Reynolds numbers with large eddy dynamics.

There are two ways to extend the LB method to include
small scale dynamics for high Reynolds number flows. First, since the
LB method was originated from lattice gas automaton method
and the lattice gas dynamics contains small scale
fluctuations intrinsically, it is natural to consider a hybrid method, in which
  the LB method is used for large scale motions and the
lattice gas  method is used to simulate small scale dynamics,
{\it i.e.}, at each point in
lattice Boltzmann space, one allows a sublattice to evolve according to
lattice gas rules\cite{brosl}.
Unfortunately, no progress has been made using this approach so far. The
fundamental difficulty there is to incorporate the interaction between
small scale dynamics and large scale dynamics using a microscopic particle
picture. A simpler approach is to combine the traditional subgrid
model with  a LB method. Some early work by Benzi
 {\em et al.}\cite{phy-report}
 and recent work at Los Alamos National Laboratory\cite{sterling2} and
by Qian {\em et al.}\cite{qian2} follow this approach. This research has helped
to
motivate our current studies.

Numerical simulation  using a subgrid model for the Navier-Stokes equations
by traditional numerical methods
is still an open and difficult problem. The basic idea of all
subgrid models is to make use of an assumption to include  the physical
effects that the unresolved motion has on the resolved fluid motion. These
models often take a simple form of eddy-viscosity models for the Reynolds
stress that serve to damp short-wavelength oscillations. Among the
simplest subgrid models is the standard Smagorinsky model\cite{smag}, which
 uses a positive eddy viscosity to represent small scale energy damping.
However, recent research by several groups\cite{mass} has
demonstrated that the eddy-viscosity acting on scales smaller than some
applied test grid could be large and positive at some locations and large
and negative at others. The standard subgrid model cannot represent these
``backscatter'' phenomena.
Recent development of dynamic subgrid models\cite{mass} represent the current
trend to include dependence of the subgrid model coefficients on local
quantities to account for these effects.

 From the numerical analysis point of view, the subgrid model must provide a
stable numerical scheme. On the other hand,  stable numerical schemes
contain some artificial dissipation and/or dispersion. One basic question to be
answered is whether
a solution using a stable numerical scheme can  represent the
 energy transfer between large and small scales effcetly as a subgrid model.

The present paper neither aims to develop a new subgrid model nor discusses
the relation between the subgrid model and numerical stability. We will
simply
demonstrate that the traditional subgrid model can be easily incorporated
into the framework of the LB method.
The scheme developed here can be easily extended to include the dynamic
subgrid model and others\cite{mass}.
To illustrate how to use the lattice Boltzmann subgrid model, we will
present simulation results for
 a driven cavity flow at Reynolds numbers varying from 100 to 1 million based
on the physical viscosity.

\section{Lattice Boltzmann Method and Single-Time-Relaxation-Approximation}

In the LB method, the evolution equation for
the particle distribution function, $f_i({\bf x},t)$, can be
 written as follows,
\begin{equation}
f_i({\bf x}+ {\bf e}_i,t+1)=f_i({\bf x},t)+\Omega_i (f({\bf x},t)),
\end{equation}
where $\Omega_i=\Omega_i(f({\bf x},t))$ is a local collision operator
depending on local particle distribution $f_i$ only.
The velocity vectors are ${\bf e}_i$, where  $i
=(0,1,\cdot\cdot\cdot,b)$ and $b$ is the number of  the nearest
neighbors. $b$ equals to 6 for a hexagonal lattice and 8 for a square
lattice.
The above lattice Boltzmann equation can be obtained from two different
ways: First, the lattice Boltzmann equation can be naturally derived from
the lattice gas equation for particle occupation, $N_i$, by assuming
no particle-particle correlations, where $N_i$ is a
Boolean variable. Although equation (1) has been written down earlier
by several people\cite{fhp1,wolf} to derive the asymptotic
behavior of the lattice gas, it was first
proposed by McNamara and Zanetti\cite{macn-zan} to use (1)
as a direct numerical method,
where they simply replace $N_i$ by $f_i$ in the lattice gas automaton
equation\cite{fhp1} without
changing the collision operator and streaming steps.
This produces: $\langle N_i \rangle = f_i$ and
 $\langle  \Omega_i(N({\bf x},t)) \rangle = \Omega_i(f({\bf x},t)) $, where
$\langle  \rangle$ denotes an ensemble average.
 Second, if one starts from a continuum kinetic equation for particle
distribution function, $f_i$,
\begin{equation}
 {\partial f_i \over \partial t} + {\bf e}_i \cdot \nabla f_i
 = \Omega_i,
\end{equation}
then the lattice Boltzmann equation in (1) can be regarded as a Lagrangian
solution using an Euler time step in conjunction with an upwind spatial
discretization\cite{sterling}:
\begin{equation}
\frac{f_i({\bf x},t+\Delta t)-f_i({\bf x},t)}{\Delta t}+\frac{f_i({\bf x}+%
{\bf e}_i\Delta x,t+\Delta t)-f_i({\bf x},t+\Delta t)}{\Delta x} =
\Omega_i.
\end{equation}
Upon setting the time spacing equal to grid spacing (particle with lattice
speed),
equation (3) becomes the same as (1).

An important refinement of the LB method
proposed by Higuera and Jimenez \cite{hig-jim} was to replace
the ``exact''  collision operator by a linearized collision operator.
A further simplification for the collision operator
was  offered nearly simultaneously by two groups\cite{ccmm,qian,ccm}. They
assumed
that the exact collision operator can be discarded,
provided one adopts a collision operator that leads, in a
controllable fashion, to a desired local equilibrium state. They
 chose an equilibrium distribution function that  depends only
upon the local fluid variables, which can be computed
from the actual values of the local distribution at a point and which
leads to the desired macroscopic equations.

For developing a magnetohydrodynamic (MHD) lattice Boltzmann
 method, Chen {\it et al.}\cite{ccmm} offered the first suggestion that one
could further
simplify the collision operator by using a single-time-relaxation
approximation, or STRA.
Subsequently, a similar method\cite{qian} was described, and referred
to as a ``BGK'' collision integral, in reference to the
more elaborate collision treatment of Bhatnagar, Gross and Krook\cite{BGK}.
The essence of the suggestion for the LB method is that
the collision term, $\Omega(f)$, be
replaced
by the well-known single-time-relaxation approximation,
$\Omega(f) = -\frac{f-f^{eq}}{\tau}$.
The appropriately chosen equilibrium distribution is denoted by
$f^{eq}$ and  depends upon the local fluid variables. A lattice relaxation
time, $\tau$,
controls the rate of approach to equilibrium. Later, Qian {\it et
al.}\cite{qian} and Chen {\it et al.}
\cite{ccm} described an STRA method for hydrodynamics that incorporates
a rest particle state in order to be consistent with the exact Navier-Stokes
equations.

To derive the Navier-Stokes equations, the Chapman-Enskog procedure is
utilized, assuming the following multi-scale
expansion of the time and space derivatives in the small parameter,
$\epsilon $:
\begin{eqnarray}
\frac \partial {\partial t}= \epsilon \frac \partial {\partial t_1}+
\epsilon^2 \frac
\partial {\partial t_2}+..., \nonumber \\
\nabla = \epsilon\nabla _1+\epsilon^2 \nabla _2+\cdot\cdot\cdot.
\end{eqnarray}

We also expand the distribution function as
\begin{equation}
f_i=f_i^{(0)}+\epsilon f_i^{(1)}+\epsilon ^2f_i^{(2)}+\cdot\cdot\cdot,
\end{equation}
where the zeroth-order term is the equilibrium distribution function, so that
the collision operator becomes
\begin{equation}
-\frac{(f_i-f_i^{eq})}{\tau}=-\frac 1\tau
(f_i^{(1)}+\epsilon f_i^{(2)}+...).
\end{equation}
Note that the parameter $\epsilon$ can be regarded as a Knudsen number
similar to kinetic theory in classical statistical mechanics.
Since the mass, $n$, and the momentum, $n{\bf u}$, are conserved in collisions:
\[ n = \sum_i f_i =
\sum_i f_i^{(0)}, n{\bf u} = \sum_i f_i{\bf e}_i = \sum_i f_i^{(0)}{\bf e}_i,\]
the summations over nonequilibrium populations are zero:
$\sum_if_i^{(l)}=0$ and $\sum_i{\bf e}_if_i^{(l)}=0$ for $l>0$.

Substituting the above expansions into (1), we obtain
equations of first and second order in $\epsilon $ which are written as
\begin{eqnarray}
\frac{\partial}{\partial t_{1}}f^{(0)}_{i}+{\bf e}_{i} \cdot {\bf \nabla}_{1}
f^{(0)}_{i} = -\frac{1}{\tau}
f^{(1)}_{i},
\end{eqnarray}
and
\begin{equation}
{\partial f_i^{(0)} \over \partial t_2} + ({\partial\over \partial t_1} +
{\bf e}_i \cdot {\nabla}_1) (1 - \frac{1}{2\tau}) f_i^{(1)}
= - \frac{f_i^{(2)}}{\tau}.
\end{equation}
where $\tau$  is assumed to be $O(1)$.

When equations (7) and (8) are summed over the $i$ velocities, the
continuity or mass conservation equation to second order in $\epsilon $ is

\begin{equation}
\frac{\partial n }{\partial t} +
{\bf \nabla}\cdot (n{\bf u}) = 0.
\end{equation}
The momentum equation to second order in $\epsilon $ is obtained by
multiplying the above equations by ${\bf e}_i$ and then summing over
velocities,
\begin{eqnarray}
\frac{\partial }{\partial t}(n{\bf u})
+{\bf \nabla}\cdot ({\bf \Pi}^{(0)}+ (1 - \frac{1}{2\tau}){\bf \Pi}^{(1)}) = 0,
\end{eqnarray}
where ${\bf \Pi }^{(0)}$ and ${\bf \Pi }^{(1)}$ are the momentum flux tensors,
 defined as
\begin{equation}
{\bf \Pi }_{\alpha \beta }^{(0)}=\sum_ie_{i\alpha }e_{i\beta }f_i^{(0)},
\end{equation}
and
\begin{equation}
{\bf \Pi }_{\alpha \beta }^{(1)}=\sum_ie_{i\alpha }e_{i\beta }f_i^{(1)}.
\end{equation}
Note that the factor $\frac{1}{2\tau}$ comes from the second time and space
derivatives\cite{schen4}.

The constitutive relations for this tensor are obtained by
selecting a particular lattice geometry and equilibrium distribution
functional form and then proceeding to match moments of the distribution
function with terms in the Navier-Stokes equations.

As an example, we use in this paper a 9-velocity
square lattice with velocity vectors, ${\bf
e}_{i}^{I}=\{cos(\pi(i-1)/2),sin(\pi(i-1)/2)\}$ and ${\bf
e}_{i}^{II}=\sqrt{2}\{cos(\pi(i-\frac{1}{2})/2),sin(\pi(i-\frac{1}{2})/2)\}$
for $i=1, \cdot\cdot\cdot,4$ .
A suitable equilibrium distribution function
is found to be
\begin{eqnarray}
f_{0}^{\rm eq} =\frac{4}{9}n [1-\frac{3}{2}u^{2}], \nonumber \\
f_{i}^{I,\rm eq} = \frac{n}{9}[1 + 3{\bf e}_{i}^{I}\cdot{\bf u}
+ \frac{9}{2}({\bf e}_{i}^{I}\cdot{\bf u})^{2}
-\frac{3}{2} u^{2}],\nonumber \\
f_{i}^{II,\rm eq} = \frac{n}{36}[1 + 3{\bf e}_{i}^{II}\cdot{\bf u}
+ \frac{9}{2}({\bf e}_{i}^{II}\cdot{\bf u})^{2}].
-\frac{3}{2} u^{2}.
\end{eqnarray}

Substituting equation (13) into equation (10) for ${\bf \Pi }$
above and noting the velocity moment relations on the square
lattice\cite{schen4},
we find that
\[ {\bf \Pi }_{\alpha \beta }^{(0)}=\frac{n}{3}\delta _{\alpha \beta
}+nu_\alpha u_\beta.  \]
This provides a Galilean invariant convective term in the momentum equation.
By identifying the isotropic part of this tensor with the pressure, we obtain
an ideal gas equation of state ({\it i.e.} $p=\frac{n}{3}$).

Neglecting terms of order $u^3$ and higher, we have:
 \[
{\bf \Pi }_{\alpha \beta }^{(1)}=-(\tau-\frac{1}{2})\{\frac \partial
{\partial t}{{\bf %
\Pi }_{\alpha \beta }^{(0)}}+\frac \partial {\partial x_\gamma
}\sum_ie_{i\alpha }e_{i\beta }e_{i\gamma }f_i^{(0)}\}. \]

{}From the distribution function in (13)\cite{hou}, we have:

\begin{equation}
{\bf \Pi }_{\alpha \beta }^{(1)}=-\frac{2n\tau}{3}S_{\alpha\beta},
\end{equation}
with $S_{\alpha\beta} = \frac{1}{2}(\frac{\partial u_\beta }
{\partial x_\alpha }+\frac{\partial u_\alpha }{%
\partial x_\beta })$.

Upon substitution into equation (10), the final form of the momentum
equation is
\begin{equation}
n\frac{\partial u_\alpha }{\partial t}+nu_\beta \frac{\partial u_\alpha }{%
\partial x_\beta }=-\frac{\partial p}{\partial x_\alpha }+\frac \partial
{\partial x_\beta }(\mu (%
\frac{\partial u_\beta }{\partial x_\alpha }+\frac{\partial u_\alpha }{%
\partial x_\beta })),
\end{equation}
where $\mu$ is the shear viscosity,
\begin{equation}
\mu =\frac{(2\tau - 1)n}{6},
\end{equation}
and this gives a kinematic viscosity of $\nu = \frac{(2\tau - 1)}{6}$.

There are differences between the incompressible Navier-Stokes equations and
the macroscopic behavior of the discrete-velocity Boltzmann equations derived
above
because of the asymptotic nature of the Chapman-Enskog method. The
differences may be attributed to two causes: 1) Burnett level and higher level
terms or as
small deviations from the above relation for the kinematic viscosity;
2) high order velocity terms\cite{qian3} and compressibility effects.
Since the Knudsen number is proportional to the Mach number
divided by the Reynolds number, the Burnett terms may be classified with
other ``compressibility'' effects and should become small as the Mach number
approaches zero for a fixed Reynolds number.

The particle interpretation of the LB method allows boundary
conditions to be implemented as particular types of collisions.
If populations are reflected directly back
along the lattice vector from which they came, the result is a
``no-slip'' velocity boundary condition. One may also define specular
reflection conditions that yield a slip condition. These simple
boundary conditions make the LB method particularly suited to parallel
computing environments and the simulation of flows in complex geometries.
It should be mentioned, however, that the recent work by Hou
{\em et al.}\cite{hou}has
 demonstrated that even though the LB method discussed above is
of second order accuracy for both space and time discretization, the simple
particle bounce back leads to a first order accuracy in spatial discretization
at the boundary.
Nevertheless, second order schemes similar to bounce back have been
proposed recently by Skordos\cite{sko} and Noble {\em et al.}\cite{nobel}.

Since the LB method under consideration is valid only in the incompressible
limit, the main dimensionless parameter of interest is the Reynolds number.
Here the Reynolds number is defined as,
\[ Re=\frac{LU}\nu =\frac{6N U}{\tau -\frac{1}{2}}, \]
where $N=\frac L{\Delta x}$ is the number of lattice spaces and $U$ is a
characteristic velocity. Convergence of the solution to the
incompressible Navier-Stokes equations
for a fixed Reynolds number is then obtained by letting the Mach number
become small enough to remove compressibility effects, and letting the
lattice spacing ${\bf e}_i\Delta t$ become small enough to resolve the
flow. Several recent studies have shown\cite{hou,rei,daniel} that the lattice
Boltzmann
methods accurately predict incompressible fluid flows in this limit.
Moreover, the convergence at a fixed
Reynolds number is performed by increasing $N$ while either increasing $
\tau $ and/or decreasing $U$ appropriately. For a decrease in the value
of $U$, a proportionate increase in the number of time steps is needed
to reach the same flow evolution time.

\section{The Lattice Boltzmann Subgrid Model}

To simulate large scale resolved fluid problems at high Reynolds numbers,
a space filtering operation is often introduced,
\begin{equation}
\label{eq:filter}\overline{w}\left( x\right) =\int w\left( x\right) G\left(
x,x^{\prime }\right) dx^{\prime },
\end{equation}
where $w$ can be density, velocity or any other physical quantity;
$G$ is a given spatial filter function and the integral is extended over the
entire domain. Several different filters\cite{fer} can be used, depending
 on the numerical method in use. For most finite difference methods, the box
filter is
assumed and is defined as follows:
\begin{equation}
\label{eq:filter1}G_i\left( x_i,x_i^{\prime }\right) =
\begin{array}{c}
\frac 1{\Delta _i}
{ for }\left| x_i-x_i^{\prime }\right| <\frac{\Delta _i}2, \\ 0{
otherwise,}
\end{array}
\end{equation}
while for spectral methods, a cutoff filter defined in Fourier space is used,
\begin{equation}
\label{eq:filter2}\widehat{G}_i\left( k_i\right) =
\begin{array}{c}
1
{ for }k_i<K_i \\ 0{ otherwise,}
\end{array}
\end{equation}
where $\widehat{G}_i$ is the Fourier coefficient of the filter function in
the ith direction, $G_i$, $K_i=\pi /\Delta _i$ is the cutoff wavenumber, and
$\Delta _i$ is the filter width in the ith direction.
Applying the same filtered operation to density, pressure and velocity in
equation (9) and (15) in the incompressible limit, it can be easily shown that
 the filtered continuity and momentum equations are
\begin{equation}
\label{eq:Navier-Stokes}
\begin{array}{c}
\frac{\partial \overline{u}_i}{\partial x_i}=0 \\ \frac{\partial \overline{u}%
_i}{\partial t}+\overline{u}_j\frac{\partial \overline{u}_i}{\partial x_j}%
=-\frac 1n\frac{\partial \overline{p}}{\partial x_i}-\frac{\partial \tau
_{ij}}{\partial x_j}+\frac \partial {\partial x_j}\left( \nu \left[ \frac{%
\partial \overline{u}_i}{\partial x_j}+\frac{\partial \overline{u}_j}{%
\partial x_i}\right] \right),
\end{array}
\end{equation}
where $\tau_{ij}$ is the Reynolds stress,
representing the effects of the unresolved scales on the resolved scales:
\begin{equation}
\label{eq:Reynoldsstress}\tau _{ij}=\overline{u_iu_j}-\overline{u}_i%
\overline{u}_j
\end{equation}
These equations govern the evolution of the resolved fluid motions. For
laminar flows, today's computational power is often adequate to resolve
all scales, which in the context of filtering simply means that $\tau _{ij}$
is negligible. For turbulent flows, however, the
Reynolds stress is significant and must be modeled in some way.
The use of artificial dissipation or dispersion in any numerical schemes to
stablize the solution
is equivalent to setting discretization error equal to this Reynolds stress
term.
Thus, any numerical viscosity will effectively serve as a subgrid
turbulence model. The main criterion for stability of numerical methods is
that this term effectively damps short-wavelength oscillations
in the flow.

The most common approach to subgrid modeling is due to Smagorinsky
\cite{smag} in which the anisotropic part of the Reynolds stress term
is modeled as
\begin{equation}
\label{eq:smagmodel}\tau _{ij}-\frac{\delta _{ij}}3\tau _{kk}=-2\nu _t%
\overline{S}_{ij}=-2C\Delta ^2\left| \overline{S}\right| \overline{S}_{ij},
\end{equation}
in which $\delta _{ij}$ is the Kronecker delta function
and $\left| \overline{S}%
\right| =\sqrt{2\overline{S}_{ij}\overline{S}_{ij}}$ is the magnitude of the
large scale strain rate tensor
\begin{equation}
\label{eq:strainrate}\overline{S}_{ij}=\frac 12\left( \frac{\partial
\overline{u}_i}{\partial x_j}+\frac{\partial \overline{u}_j}{\partial x_i}%
\right) .
\end{equation}
and $C > 0$ is the Smagorinsky constant. The isotropic part of the
Reynolds stress term can be included in the pressure term.

A recent modification to this model by Germano {\em et al.}
\cite{mass} is called the dynamic subgrid eddy viscosity model and applies
a test filter in addition to the grid filter. The so-called ``resolved
Reynolds stress'' terms that are computed for scales between the coarse test
filter and the grid filter can then be used to locally compute the
Smagorinsky constant, $C$. A problem with this model, however, is that the
Smagorinsky constant may become locally negative and numerical instability
ensues. This is avoided in practice by averaging over homogeneous flow
planes to keep positive eddy viscosities or by developing even more
sophisticated
and more complicated models.

The fundamental question we are facing is how to modify the lattice Boltzmann
method discussed in the previous section to
simulate the filtered density and velocity in equation (15). In order to
apply the subgrid idea in the framework of LB method,
let us introduce the filtered particle distribution, ${\overline f_i}$,
defined as follows:
\[ \overline{f_i}\left( x\right) =\int f_i\left( x\right) G\left(
x,x^{\prime }\right) dx^{\prime },  \]
and modify the microscopic equation (2) to be a kinetic equation for
the filtered particle distribution function:
\begin{equation}
 { \partial {\overline f_i} \over \partial t} + {\bf e}_i \cdot \nabla
{\overline f_i}
 = {\overline \Omega_i}.
\end{equation}

Using the finite-difference scheme in (3) for (24), we will obtain similar
equation in (1) for ${\overline f_i}$.

There are two possible
approaches. First, one can use the unfiltered equilibrium distribution (1)
as the  fundamental equation. When we apply the $G$-filter on the
collision term, or the equilibrium distribution function, we obtain
several correlation terms due to the nonlinearity of velocity in
the equilibrium distribution function: ${\overline {nu_{\alpha}}}$
 and ${\overline {nu_{\alpha}u_{\beta}}}$,  which need to be closed.
Therefore, the closure problem is again encountered. It can be argued that
these particle-particle space correlations may be easier to model
using physical arguments based on
 microscopic simplicity. The final results of modeling these terms
should be equivalent to modeling the Reynolds stress at the macroscopic level.

A second and more direct approach is to link the
collision steps with some local information and abandon the
single-time-relaxation approximation. To do this, we assume that the
filtered particle distribution will approach a local filtered
equilibrium distribution, which can be chosen to depend only
 on local filtered
mean quantities, ${\overline n}$ and ${\overline {\bf u}}$, {\it i.e.},
${\overline {\Omega_i (f({\bf x},t))}} =
\Omega_i ({\overline {f({\bf x},t)}})$ and the form of
the equilibrium distribution function are all the same as in (13),
except now we will use the filtered quantities to replace the unfiltered
quantities. In addition, we further assume that the effect of the
correlation will only locally introduce an eddy viscosity and we allow
the relaxation time to  depend on space. Explicitly, we will incorporate
the Smagorinsky formula for eddy viscosity into the relaxation time, $\tau$.
Since in the LB method, the lattice spacing is set to 1, the relaxation
time is equivalent to the mean free path of a particle. The above arguments
can be directly linked to the
early idea in  Prandtl's mixing length theory\cite{les}, {\it i.e.} the change
of
relaxation time can be equivalently interpreted as a change in the local
mean free path and therefore the local viscosity.

It can be easily proved that the spatial dependence of the
relaxation time, $\tau$, will
not change the Chapman-Enskog expansion procedure and
 does not affect the derivation of the Navier-Stokes equations
in the last section. That means equations
(9) and (15) can also be derived using equation (24), assuming:
${\overline n} =  \sum_i {\overline f_i}$ and ${\overline n}
{\overline {\bf u}}=  \sum_i {\overline f_i}{\bf e}_i$.
We also find that if the total viscosity equals the sum
of the physical and the eddy viscosities (the Smagorinsky model),
\begin{equation}
\nu _{total}=\nu_0 + C\Delta ^2\left|
\overline{S}\right| ,
\end{equation}
where $\nu _{total}$ is the total viscosity; $\nu_0$ is the physical
kinematic viscosity and $C\Delta ^2\left|
\overline{S}\right|$ is the eddy-viscosity term, we should let the
 relaxation time, $\tau$, be a function of total viscosity.  Using (16), we
have,
\[ \frac{2\tau_{total} -1}{6} = \nu _{total}, \]
and we obtain a simple relation:
\[ \tau_{total} = 3\nu _{total} + \frac{1}{2}, \]
or
\[ \tau_{total} = 3(\nu_0 +  C\Delta ^2\left|
\overline{S}\right|) + \frac{1}{2}.  \]
Thus, the value of $\tau$ should be locally adjusted, depending on the local
magnitude of the large scale strain rate tensor.  The dynamic eddy viscosity
model may also be implemented by using the local value of $C$ that is
computed from the resolved Reynolds stress term using the test
filter\cite{mass}.

It should be mentioned that the calculation of local intensity of the
strain tensor can be easily carried out using the nonequilibrium properties
of the filtered particle distribution. The procedure is:

\noindent
(a) calculate locally the nonequilibrium stress tensor:
\[ {\overline \Pi_{i,j}} = \sum_{\alpha} {{\bf e}_{\alpha}}_i{{\bf
e}_{\alpha}}_j ({\overline f_i} -
{\overline f_i^{eq}}); \]

\noindent
(b) calculate the second variance of the tensor $\Pi_{i,j}$,
Q = ${\overline \Pi_{i,j}}{\overline \Pi_{i,j}}$;

\noindent
(c) neglecting high order velocity effects and using equation (14),
we prove that the following relation is  valid:
\begin{equation}
Q^{1/2} = (\tau_0 + 3C \Delta^2 |{\overline S}|)\frac{2n}{3}|{\overline S}|,
\end{equation}
where $|\overline{S}|$ is the intensity of the local filtered strain tensor
and $\tau_0 = 3\nu_0 + \frac{1}{2}$. Solving the above equation for
$|\overline{S}|$, we have:
\[
|{\overline S}| = \frac{\sqrt{\nu_0^2 + 18 C \Delta^2 Q^{1/2}} -
\nu_0}{6C\Delta^2}. \]
The solution through the above equation guarantees  the required
positivity of
$|{\overline S}|$.

In a traditional scheme, a finite difference must be used for calculating the
local strain tensor. In the present model, the strain tensor can be
calculated locally using the nonequilibrium distribution. This simple
procedure should save considerable computational time.

\section{Turbulence Structures in a Driven Cavity Using a Lattice Boltzmann
Simulation}

  The two-dimensional driven cavity flow considered here is used as a testbed
for the subgrid model described above. The incompressible fluid is bounded in
a square enclosure. The flow is driven by the uniform translation of the
top boundary. The flow configurations generated in this cavity show rich vortex
phenomena at many scales depending on the Reynolds number, Re. Cavity flow
has been used extensively in various numerical schemes as an ideal problem  for
studying complex flow physics in a simple geometry.

  Most numerical solutions of two-dimensional cavity flow use a
vorticity-stream function formulation and discretize the incompressible,
steady linear or nonlinear Navier-Stokes equations by finite difference,
finite element methods or their combinations with multigrid and other methods.
Among the previous research, Ghia {\it et al.} \cite{ghia} obtained numerical
solutions up to Re=10,000 with a 257$ \times $257 grid. Their work represents
the most comprehensive study of 2-D cavity flow to date.

The cavity flow using the lattice Boltzmann method without subgrid model are
studied carefully by Hou {\em et al.}\cite{hou}. Detailed quantitative
comparisons between
the LB and traditional methods for laminar flow (Re $\le $ 7500) were carried
out. The compressibility error and the convergence rate of the
method, as well as parameter ranges for stable simulations are also explored
in that paper. The results obtained by LB method  compared well with
Ghia {\it et al}. \cite{ghia}.  The maximum and minimum values of stream
function for primary and secondary vortices agree
with each other within 0.2 \% for all values of Re tested.
The locations of the vortex centers predicted by the lattice Boltzmann method
also agree well with those given by Ghia {\it et al}. \cite{ghia}.

Most previous finite-difference or finite-element methods are only concentrated
in steady-state behavior and therefore only the
steady-state, partial differential equations are used. In contrast, the LB
method is a direct numerical scheme based on the unsteady microscopic
kinetic equation. The time dependent behavior will be calculated in spite of
the time-dependent properties of the flow. In some sense,
the LB method is close to a time-dependent method in traditional finite
schemes. It is quite encouraging that even though
 the LB method is very different from
other traditional methods from lots of aspects, including
microscopic vs. macorscopic and time-dependent vs. time-independent,
 the dynamic properties of the flow agree each other surprisingly well.
Nevertheless, it should be realized that
the highest Re number for a cavity flow is limited about 10,000 using a
LB method without subgrid model on a 256 $\times $ 256 lattice. If viscosity is
decreased further to increase Re, the computational
instability will cause the simulation to blow up.  At such a high Re,
flows are turbulent. The
value of Re can also be increased by increasing the number of lattice sites and
increasing the maximum velocity. The former is limited by computer memory and
the
latter increases the compressibility error. It appears that the original
lattice Boltzmann model is more adequate to laminar flow with
low Reynolds numebr
than to  turbulent flow with high Reynolds number. The combination of the LB
method with the subgrid model
described in this paper allows simulations of cavity flow to reach up to a
Reynolds
number of $10^6$.

  The present simulation uses Cartesian coordinates with the origin located
at lower left corner. The top wall moves from left to right with a uniform
velocity,
$U$. The cavity has 256 lattice units on each side. Initially, the velocities
at all nodes, except the top, are set to zero. The x-velocity of the top is
$U$ and the y-velocity is zero. A uniform initial particle density is imposed
such that the moving particle along the
${\bf e}^{I}_i$ direction has a density fraction of $d=\frac{{\overline n}}{9}
=0.3$ per direction. The moving particle along the ${\bf e}_{i}^{II}$ direction
 has a density fraction of
$\frac{d}{4}$, and the rest particle has a density of $4d$.
Hence, the total density per node is ${\overline n} =2.7$. Using the uniform
density distribution and velocities given above, the equilibrium particle
distribution function, ${\overline f_i}$, is calculated according to (13). The
evolution of ${\overline f_i}$
can then be found by a succession of streaming and
collision-like processes. After streaming, the velocity of the top lid is
reset to the uniform initial velocity. After the streaming and
collision process cycles, the particle distribution function, ${\overline
f_i}$, at the top
is set to the equilibrium state and bounce-back boundary conditions are
used at the three stationary walls. The two upper corners are singular points
which are considered as part of the moving lid in the simulations, but tests
shown there is little difference if these two points are treated as fixed wall
points. The uniform velocity of the top wall used in the simulations was
$U$=0.1. The Reynolds number used in the cavity simulation is defined as
$Re=$U$ L/\nu$, where $U$ is the uniform velocity of the top plate, $L$ is the
edge length of the cavity and $\nu$ is the kinematic viscosity related
to the single relaxation time as given in (16). All results are
normalized to allow comparisons between the present work and other
results based on a unit square cavity with unit velocity of the top
boundary\cite{ghia}.

  Steady-state solutions for cavity flow are obtained using the lattice
Boltzmann method with subgrid model for Re=100, 2,000, and 7,500.
The Re=10,000 case is also run on a 256$ \times $256 lattice, but steady state
cannot be reached because bifurcation takes place somewhere between Re=7,500
and 10,000. The results for Re=10,000 oscillate between a series of different
configurations. All of the results obtained by the present model for laminar
flow ($Re < 7500$) agree well with that obtained
 by LB method without using subgrid model.
Three Smagorinsky constants, {\it i.e.}, $C=0.0025, C=0.01,$ and
$ C=0.04$ are
tested for these simulations. The two former cases give relatively smaller
error for the velocity field. For $Re=10^5$ and $Re=10^6$ , the Smagorinsky
constant we are using is higher than the value quoted from other traditional
schemes\cite{mass} and it cannot be lower
than 0.073 and 0.084, respectively, otherwise the
particle distribution function becomes negative, which means the eddy
viscosity is not sufficient to damp the high-frequency fluctuations at small
scales. It appears that the value of the Smagorinsky constant stongly depends
on the geometry.

  Figure 1 shows plots of the stream function for the Reynolds numbers
considered. It is apparent that the flow structure is in good agreement with
Ghia {\it et al}.  for laminar flow. These plots give a clear
picture of the overall flow pattern and the effect of
Reynolds number on the structure of the recirculating eddies in the
cavity. In addition to the central vortex, a pair of counterrotating
eddies of much smaller strength develop in the lower corners of the cavity at
Re=100. At Re=2000, a third secondary vortex is observed in the upper
left corner. For Re $\ge $5,000, a tertiary vortex in the lower right hand
corner appears. For $Re=10^5$ and $10^6$, a series of vortices are
formed at the lower boundary. For low Re ({\it e.g.} Re=10,
not shown here), the center of
the primary vortex is located at the midwidth and at about one third of the
cavity depth from the top. As Re increases ($Re  =  100$), the primary vortex
center moves towards the right and  becomes increasing circular. This center
moves down towards the geometric center of the cavity as the Re
increases and becomes fixed in its x location for Re $\ge$ 5,000.
For turbulent flow the primary vortex is no longer round.  In the
case of $Re=10^5$,
the core of the primary vortex
changes its shape and rotates with time. When the value of Re as high as
$10^6$, the vortex street is formed inside the primary vortex.

  The plots of vorticity in Figure 2 show that the laminar cavity flow
within closed streamlines at relative high Re consists of a central, inviscid
core of nearly constant vorticity with viscous effects confined to thin shear
layers near the walls. As Re increases, several regions of high vorticity
gradients (indicated by concentration and wiggle of the vorticity contours)
appear within the cavity. However, for turbulent flow there is no such inviscid
core and a full developed turbulence is seen at $Re=10^6$.
Quantitative study of turbulence by present model is undertaken. Improvement of
the subgrid model used in the lattice Boltzmann method is also needed.

\section{Concluding Remarks}

In this paper, we have presented a lattice Boltzmann subgrid model for
simulating
fluid flows at high Reynolds numbers. The essential idea is to define
a space-filtered particle distribution
and to allow the dynamics of the filtered particle distribution to have a
space-dependent relaxation. This is equivalent to Prandtl's mixing length
theory for which the mean free path of particle will be affected by the local
strain
intensity. The current model is able to incorporate the standard Smagorinsky
subgrid model to include the energy dissipation induced by
the interaction between resolved and unresolved scales. The lattice Boltzmann
method proposed in this paper does not require calculation of the local
strain intensity (nor the Reynolds stress tensor) using a finite difference
scheme.
Instead, the Reynolds stress is calculated
locally at each time step using the nonequilibrium particle distribution
functions. This additional advantage may allow the computational program
to be more efficient than traditional finite different schemes.

The application of the current lattice  Boltzmann model to a driven cavity
fluid
flows over a large range of Reynolds numbers has been carried out.
The simulation results agree well with other methods for low Reynolds number
fluid flows and qualitatively agree with previous simulations for high Reynolds
numbers.

There are several issues to be addressed. First, to stablize numerical
simulations, the current lattice Boltzmann subgrid model
requires the Smagorinsky constant slightly bigger than the value commonly used
in other
geometries. This could be an indication of the geometry-dependence of the
constant as
observed in other finite difference simulations. It could also be attributed to
the
fact that the current lattice Boltzmann method is only accurate to first order
at
the boundaries\cite{hou}.

Second, it should kept in mind that the development of the lattice Boltzmann
subgrid model is necessitated  by a numerical instability encountered in
lattice Boltzmann simulations at  high Reynolds numbers. As mentioned in
Section 1,
the lattice gas automaton  is unconditionally stable. Using real number
manipulation
in the LB method does reduce the noise and possibly utilizes memory more
efficiently\cite{schen4}.
The real number operations introduce the same numerical round-off error
 and, therefore, instability problem encountered in other explicit finite
different schemes.
Moreover, the introduction of {\em ad hoc} assumptions in subgrid models can
not be theoretically
justified at the present time. The current subgrid  models must be considered
to be phenomenological.  It is desirable that lattice gas models could
 simulate fluid flows with high Reynolds number.
 Unfortunately all known lattice gas models can only simulate relatively
large viscosities, {\it i.e.} low Reynolds number. At the moment, it is not
clear if the lattice Boltzmann subgrid model is better or worse than lattice
gas
model or other traditional finite difference schemes for simulating
high Reynolds number flows.

Third,  it appears that more careful
quantitative comparisons between the lattice Boltzmann subgrid
 method and other numerical schemes are needed. These comparisons should
include
studies of numerical accuracy and computational speed. In addition, the current
simulation only involves a simple geometry.
It would be interesting to apply the current scheme to complicated geometries
and to
complicated physical phenomena, such as the multiphase fluid flows, for
which the traditional schemes may have difficulties.

As a final remark, we emphasize that the current scheme can be
extended easily to  other dynamic subgrid models. Both the standard
Smagorinsky model and the dynamic model of Germano {\it et al.} \cite{mass}
have been implemented on the CM-5 computer at the Los Alamos Advanced
Computing Laboratory to understand fluid instabilities and high Reynolds
number flows\cite{sterling2}. It also should be pointed out that
other possible subgrid models at the particle level, such as the model
based on the first approach in the discussion of Section 3 may lead to more
physically based turbulent models and  could link the microscopic world
with the macroscopic effective subgrid viscosity.

\section{Acknowledgements}

We thank Y. H. Qian for useful discussions.
This work is supported by the U.S. Department of Energy at
Los Alamos National Laboratory. Numerical
simulations were carried out on the CM-5 at the Advanced Computing Laboratory
at Los Alamos National Laboratory.

\vfill\eject

\section{Figure Captions}
\begin{description}

\item[Fig. 1]Contour plots of the stream function at Re = 100, 2000, 7500,
$10^4$,
$10^5$ and $10^6$.

\item[Fig. 2]Contour plots of the vorticity distribution function at Re = 100,
2000, 7500, $10^4$,
$10^5$ and $10^6$.

\end{description}

\end{document}